# THE N-BOX PARADOX IN ORTHODOX QUANTUM MECHANICS

By CONALL BOYLE [*]
University of Central England, Birmingham B42 2SU, U.K.

and ROGER SCHAFIR [†]
CISM, London Guildhall University, London EC3N 1JY, U.K.

ABSTRACT

The prediction of the N-box paradox, that whichever box is opened will contain the record of the particle having passed through it, is traced to a failure to specify whether the other boxes are distinguishable or indistinguishable. These correspond to different ways of lifting the degeneracy of a certain measurement, and have incompatible consequences.

There is now a considerable amount of literature on the N-box paradox [1-9], much of it concerned with interpretational matters, such as its role in rival interpretations of quantum mechanics or the correct use of counterfactuals. These are interesting, but the startling aspect of the matter is that *orthodox* quantum mechanics apparently makes an impossible prediction at the *observational* level of macroscopic outcomes and observer's choices of what to measure. This should be headline news if true. But is it really true?

First let us summarise the N-box paradox. A quantum system (a 'particle') with an $N+1$ dimensional state space starts off in the prescribed state:

$$|initial> \;=\; \frac{1}{\sqrt{N+1}} \sum_{j=1}^{N+1} |j>, \qquad (1)$$

and ends in the prescribed state:

$$|final> \;=\; \frac{1}{\sqrt{N^2-N+1}} \left\{ \sum_{j=1}^{N} |j> - (N-1)|N+1> \right\}. \qquad (2)$$

Then if during the intermediate passage the measurement was made which is described by the pair of projection operators:

$$|i><i|, \;\; I - |i><i|, \qquad (3)$$

for any single choice of $i$ except $N+1$, it will be found with certainty that the particle was projected into the state $|i>$.

So, thinking of the basis states $|i>$ as 'boxes', the N-box paradox is stated like this: 'The particle is found with certainty in any one of N boxes.'

We recognise of course that we only look at the sub-ensemble for which the particle is projected into the final state (2), and that the intermediate measurement must already have been made, and has been recorded — or at least must be regarded as recordable — by the time the particle reaches the final state. It need not be paradoxical that the effect on the particle, if one box is examined and no others, is that it can only be projected into the final state if it was in the examined box. Nevertheless, if we imagine, for example, that the observer of the final state is different to the observer who opens the box, and that the choice of box is kept secret from the final observer, who is then, however, asked to guess which box was opened, we seem to have a prediction that whichever guess the final observer makes is bound to be correct. This is because the observables $|i><i|$, $I - |i><i|$, in (3) commute with any other such pair $|j><j|$, $I - |j><j|$, for $i, j \neq N+1$, so whichever pair was actually measured, we can attribute the same assignment of values to the other pair simultaneously, on the grounds that these values would definitely have been obtained if the alternative pair had been measured instead of the actual pair which was measured, and the particle would not have been affected in its subsequent behaviour [1].

By comparison, if the 'all boxes are opened' measurement had been made at the intermediate time, i.e. the projectors:

$$|i><i|, \text{ for } i = 1,2,...,N+1, \qquad (4)$$

there would be an equal probability for each $i$ of the particle being found in Box $i$.

To proceed it will be helpful to compare the choice of projectors in (3) above with a certain different choice. Suppose we chose:

$$|i><i|, \left\{\sum_{j \neq i} |j>\right\}\left\{\sum_{k \neq i} <k|\right\}, [2] \quad \text{ for } i \neq N+1. \qquad (5)$$

(plus further projectors to make a complete orthonormal set; however all but the above two are orthogonal to the initial state (1) and so have zero probability of being attained). These projectors represent a measurement of whether or not the particle is in Box $i$ *when all the other boxes are indistinguishable*, since the second projector is the projection into an equally weighted superposition of all other box possibilities. This second projector in (5) has the same effect as the second projector in (3): they both project the particle into the state:

$$\frac{1}{\sqrt{N}} \sum_{j \neq i}^{N+1} |j>, \qquad (6)$$

which is orthogonal to the final state.

However the alternative projector in (3) is the sum of projectors:

$$\sum_{j \neq i} |j><j|, \qquad (7)$$

and this is consistent with the outcomes being distinguishable, or at least that they *could* be distinguished by further, compatible, measurements, since the probability of a 'yes' (i.e. eigenvalue 1) for $\sum_{j \neq i} |j><j|$ is the sum of the individual probabilities:

$$prob\left\{\sum_{j \neq i} |j><j|\right\} = tr\left\{|init><init|\sum_{j \neq i} |j><j|\right\} = \sum_{j \neq i} |<init|j>|^2. \qquad (8)$$

So we have the puzzle that the alternative outcome in the case where the individual boxes could be distinguished (and if they haven't been, it is only classical ignorance, not quantum unknowability) projects the system into the same state as the alternative outcome when all but one of the boxes are indistinguishable, thus wrongly giving the same probabilities relative to the final state. Let us look again at the assumption about the state the particle is projected into under the alternative outcome.

There is no reason for doubting the correctness of the state the particle is projected into if it has been projected by the indistinguishable boxes projector in (5). Correspondingly there is no mystery in saying that if there are, in effect, only two boxes, one small and one large, then the particle can be projected into the final state (2) only if it was in the smaller box.

But if the failure to know which box (other than Box $i$) it was projected into is only classical ignorance, then the correct description of the situation must be that the particle was projected into one of the states $|j>$, $j \neq i$, but we do not know which value of $j$. Given that the alternative box exists but is not known, we would expect that the probability of the final measurement projecting the particle into the final state is the sum of probabilities of projecting it from each box into the final state:

$$\sum_{j \neq i} |<j|final>|^2 \qquad (9)$$

which is consistent with assuming that the particle is projected, not into the pure state associated with the vector:

$$\sum_{j \neq i} (<initial|j>)|j>, \qquad (10)$$

but into the mixed state associated with the density:

$$\sum_{j \neq i} |<initial|j>|^2 |j><j|. \qquad (11)$$

---

[1] If the alternative pair did not commute with the actual pair, we could answer the paradox by pointing out [2,8] that if the alternative intermediate measurement had been made instead of the actual intermediate measurement which was made, the particle might not have been projected into the same final state (2).

[2] times a normalisation factor.

If we consider an ensemble of particles starting in the initial state (1), and subject to the distinguishable boxes measurement, we have a probability weighted mixture described by this density.

However in the N-box paradox it is assumed that the distinguishability is expressed by taking no more than the degenerate measurement (3), then applying the rule that a particle in the state $|\psi>$ is projected into the state $P|\psi>$ (times a normalisation factor), where $P$ is the projector into the degeneracy eigenspace. This has the consequence that although we used a sum-of-probabilities rule for projection *into* a degeneracy eigenspace in (8) above, we used the probability for projection from one definite state to another when it came to projection *from* the degeneracy eigenspace to another definite state. But in a case of classical ignorance the classical analysis would not distinguish between the two time directions.

The degenerate measurement (3) is evidently not sufficient to bear the interpretation put on it by the N-box paradox unless the degeneracy is lifted by co-measuring further compatible observables which distinguish between the boxes. A degenerate measurement on its own projects the system by the rule $|\psi\rangle \to P|\psi\rangle$, but a compatible co-measurement further projects $P|\psi>$ into an eigenstate (of the new observables) which lies in the degeneracy eigenspace. However there is the special case when $P|\psi>$ is one of the new eigenstates, and then the co-measurement makes no difference to the resulting state of a system that was initially in the state $|\psi>$.

The indistinguishable boxes measurement (5) is precisely such a case. Projection by $I-|i><i|$ gives the same result as projection by $\left\{\sum_{j \neq i} |j>\right\}\left\{\sum_{k \neq i} <k|\right\}$, which makes the indistinguishable boxes measurement the non-degenerate (or less degenerate) equivalent measurement to (3). In this sense the indistinguishable boxes measurement must be regarded as the correct interpretation of (3) for particles which were in the initial state (1).

The N-box paradox, then, contains an issue about a quantum state plus an element of stage magic, because its wording suggests that the other boxes are distinguishable but its conclusion only holds if the other boxes are indistinguishable. The technical point is that the degeneracy must be lifted for an unambiguous interpretation, which can be done in different ways by co-measuring different compatible observables (note that $I-|i><i|$ commutes with $\left\{\sum_{j \neq i} |j>\right\}\left\{\sum_{k \neq i} <k|\right\}$ in (5) as well as with each $|j><j|$ individually). But different compatible extensions may be incompatible with one another, as is the case for the distinguishable versus indistinguishable alternative boxes. The indistinguishable boxes case gives the same results as the degenerate measurement specified in (3), while if the other boxes are distinguishable the alternative state is not the projection by $I-|i><i|$, but an unknown state which can be described by a probability-weighted sum of the boxes $|j>$, formally the same as a mixed state.